# Transmit Antenna Selection for Spatial Multiplexing Systems: A Geometrical Approach

Hongyuan Zhang, Huaiyu Dai[*], *Member*, *IEEE*, Quan Zhou and Brian L. Hughes, *Senior Member*, *IEEE*

## Abstract

In recent years, the remarkable ability of multiple-input multiple-output (MIMO) wireless communication systems to provide spatial diversity or multiplexing gains has been clearly demonstrated. For MIMO diversity schemes, it is well known that antenna selection methods that optimize the post-processing signal-to-noise ratio can preserve the diversity order of the full MIMO system. On the other hand, the diversity order achieved by antenna selection in spatial multiplexing (SM) systems, especially those exploiting practical coding and decoding schemes, has not thus far been rigorously analyzed. In this paper, a geometrical framework is proposed to theoretically analyze the diversity order achieved by transmit antenna selection for independently encoded spatial multiplexing systems with linear and decision-feedback receivers. For a system with $N_T$ transmit and $N_R$ receive antennas, it is rigorously shown that a diversity order of $(N_T - 1)(N_R - 1)$ can be achieved for a SM system in which $L = 2$ antennas are selected from the transmit side. For $L > 2$, it is shown that the achievable diversity order is bounded below by $(N_T - L + 1)(N_R - L + 1)$. Furthermore, the same geometrical approach is used to evaluate the diversity-multiplexing tradeoff curves for spatial multiplexing systems with transmit antenna selection.

[*] The authors are with the Department of Electrical and Computer Engineering, NC State University, Raleigh, NC 27695-7511. Phone: (919) 513-0299; Fax: (919) 515-2285; Email: {hzhang, hdai, qzhou, blhughes}@ncsu.edu. This work was supported in part by the National Science Foundation under grant CCR 03-12696.

# I. Introduction

Multiple-input multiple-output (MIMO) techniques are expected to be widely employed in future wireless communications to address the ever-increasing demand for capacity. A major potential problem with MIMO is increased hardware cost due to multiple analog/RF front-ends, which has recently motivated the investigation of antenna selection techniques for MIMO systems [2]. In many scenarios, judicious antenna selection may incur little or no loss in system performance, while significantly reducing system cost.

MIMO systems can be exploited for spatial diversity (SD) or spatial multiplexing (SM) gains [7]. The majority of work on MIMO antenna selection focuses on the former, including selection combining, hybrid selection-maximum ratio combining (HS-MRC) [2][3], and antenna selection with space-time coding [4][5]. Essentially in these works, with independent and identically distributed (i.i.d.) Rayleigh fading path gains, the system error performance or outage probability can be readily analyzed through order statistics [20], and it has been shown that the diversity order of the full-size system can be maintained through the signal-to-noise ratio (SNR) maximization selection criterion.

By contrast, antenna selection for MIMO systems with multiple data streams, i.e., spatial multiplexing systems, has received less attention. The few existing analytical results generally assume capacity-achieving joint space-time coding and optimal decoding. The capacity-maximizing receive antenna selection is analyzed in [6], and shown to achieve the same diversity order as the full-size system. In [8], it is shown that the fundamental tradeoff between diversity and spatial multiplexing of the full-size system, obtained in [7], holds as well for MIMO systems with antenna selection.

In practice, the multiple streams in a SM system may be uncoded or separately encoded and sub-optimally decoded due to complexity concerns. In [1], several transmit antenna selection algorithms for SM with linear receivers are proposed, and some conjectures on the achieved diversity orders are made from numerical results. To the best of our knowledge, the exact diversity order achieved by antenna selection for practical SM systems has not been rigorously obtained. In contrast to MIMO diversity schemes, the key challenge that hinders accurate performance analysis is that selection is performed among a list of inter-dependent random quantities, which are correlated in a complex manner.

In this paper, we propose a new framework to theoretically analyze the diversity order achieved by transmit antenna selection for SM systems with independently encoded layers and linear or decision-feedback (DF) receivers (i.e., the V-BLAST structure [9]). In particular, we rigorously show that the optimal diversity order is $(N_T - 1)(N_R - 1)$ for an $N_R \times N_T$ V-BLAST system when $L = 2$ antennas are selected from the transmit side. This should be compared with the diversity order of a two-stream V-BLAST system without antenna selection, $N_R - 1$ [17]. Such a diversity gain can be tremendous for downlink high-data-rate communications, where there may be a large number of transmit antennas at the base stations but few receive antennas at the mobiles (e.g., $N_R = 2$). Furthermore, following the same geometrical approach, we give upper and lower bounds on the diversity order for general $L$, which coincide when $L = 2$. The corresponding diversity-multiplexing tradeoff curves are also derived. Generally speaking, our results confirm and generalize some of the conjectures in [1], thus verifying that transmit antenna selection can achieve high data rates and robust error performance in practical SM systems without complex coding.

The rest of the paper is organized as follows. The system model and problem formulation are given in Section II. The main ideas of our approach are illustrated in Section III, through the case of selecting $L = 2$ transmit antennas in the separately encoded SM systems with linear and DF receivers. The extension to the *L>2* scenarios is discussed in Section IV. Finally Section V concludes the paper with future directions.

## II. System Model and Problem Formulation

We consider a frequency non-selective block Rayleigh fading channel model, for which a SM system with transmit antenna selection can be expressed as:

$$\mathbf{y} = \sqrt{\frac{\rho_0}{L}} \mathbf{H}_{SL} \mathbf{s} + \mathbf{n}, \quad (1)$$

where the $N_R \times t$ matrix $\mathbf{y}$ is the received signal block; the $N_T \times t$ matrix $\mathbf{s}$ represents the transmitted signal block; $\mathbf{H}_{SL}$ contains the $L$ columns selected from the original $N_R \times N_T$ channel matrix $\mathbf{H} = [\mathbf{h}_1, \mathbf{h}_2, \cdots, \mathbf{h}_{N_T}]$, and $\mathbf{n}$ is the background noise. Both $\mathbf{H}$ and $\mathbf{n}$ are modeled with i.i.d. normalized complex Gaussian entries, while the independent transmitted signals (across the antennas) assume unit average energy per symbol per antenna. Therefore, $\rho_0$ is the average signal-to-noise ratio (SNR) per receive antenna. Throughout the paper we assume $N_T \geq L$ and $N_R \geq L$. Here *t* is the code length in each layer, which is actually irrelevant in our study, as layered one-dimensional coding can provide coding gain but not diversity gain.

In this paper, two sub-optimum receivers are considered: the linear decorrelating detector and the decorrelating decision-feedback detector [10], followed by single-user decoders. As we will see, the analysis for the latter relies heavily on the former. Furthermore, their diversity order analyses, a study at high SNR regimes, hold for linear MMSE and MMSE decision-feedback detectors as well (for SM systems with independently encoded layers considered in this paper). Given assumptions of i.i.d.

Rayleigh fading and $N_R \geq L$, the selected channel matrix $\mathbf{H}_{SL}$ has full column rank with probability one. For a decorrelating detector, a space equalizer $\mathbf{G} = \mathbf{H}_{SL}^{\dagger}$ is applied to the received signal $\mathbf{y}$ to obtain an estimate of the transmitted symbol vector:

$$\hat{\mathbf{s}} = \mathbf{G}\mathbf{y} = \sqrt{\frac{\rho_0}{L}} \mathbf{s} + \mathbf{G}\mathbf{n}, \tag{2}$$

where $\mathbf{A}^{\dagger}$ denotes the pseudo-inverse of matrix $\mathbf{A}$. Therefore $L$ equivalent independent single-input single-output (SISO) data links are formed as

$$\hat{s}_k = \sqrt{\frac{\rho_0}{L}} s_k + [\mathbf{G}]_{k*} \mathbf{n}, \ 1 \leq k \leq L, \tag{3}$$

where $[\mathbf{G}]_{k*}$ is the $k$th row in the matrix $\mathbf{G}$. The instantaneous channel capacity and error performance of these subchannels are determined by corresponding post-processing SNRs.

With transmit antenna selection, there are a total of $N_U = \binom{N_T}{L}$ possible antenna subsets, defined as $U_1 \sim U_{N_U}$ in the following manner:

$$\begin{aligned}
U_1 &= \{\mathbf{h}_1, \mathbf{h}_2, \cdots, \mathbf{h}_L\} \\
U_2 &= \{\mathbf{h}_1, \mathbf{h}_2, \cdots \mathbf{h}_{L-1}, \mathbf{h}_{L+1}\} \\
&\vdots \\
U_{N_1} &= \{\mathbf{h}_1, \mathbf{h}_{N_T-L+2}, \cdots, \mathbf{h}_{N_T}\}, \\
U_{N_1+1} &= \{\mathbf{h}_2, \mathbf{h}_3, \cdots, \mathbf{h}_{L+1}\} \\
&\vdots \\
U_{N_U} &= \{\mathbf{h}_{N_T-L+1}, \cdots, \mathbf{h}_{N_T}\}
\end{aligned} \tag{4}$$

where $N_1 = \binom{N_T-1}{L-1}$, and $U_1 \sim U_{N_1}$ contain $\mathbf{h}_1$. If the subset $U_j$ is selected, the post-processing SNR for the $k$th stream is proportional to the square of the projection height[1] $R^{(j)}_{k,span\{U_j(\bar{k})\}}$ from the $k$th column[2] $\mathbf{h}_k^{(j)}$ to the space spanned by the other $L-1$ selected column vectors $span\{U_j(\bar{k})\}$:

$$\rho_k = \frac{\rho_0}{L} R^{(j)}_{k,span\{U_j(\bar{k})\}} = \frac{\rho_0}{L} \left\|\mathbf{h}_k^{(j)}\right\|^2 \sin^2 \theta^{(j)}_{k,span\{U_j(\bar{k})\}}, \tag{5}$$

---

[1] Projection height refers to the norm of the error vector, i.e., the difference between a vector and its projection onto a subspace.
[2] Here $\mathbf{h}_k^{(j)}$ means the $k$th column vector in $U_j$, instead of in $\mathbf{H}$, while the similar denotations are used in $R^{(j)}_{k,span\{U_j(\bar{k})\}}$ and $\theta^{(j)}_{k,span\{U_j(\bar{k})\}}$.

where $\|\mathbf{h}_k^{(j)}\|$ is the norm of $\mathbf{h}_k^{(j)}$, while $\theta_{k,span\{U_j(\bar{k})\}}^{(j)}$ is the angle between $\mathbf{h}_k^{(j)}$ and its projection on $span\{U_j(\bar{k})\}$, defined as

$$\theta_{k,span\{U_j(\bar{k})\}}^{(j)} = \sin^{-1} \frac{\sqrt{R_{k,span\{U_j(\bar{k})\}}^{(j)}}}{\|\mathbf{h}_k^{(j)}\|}, \quad 0 < \theta_{k,span\{U_j(\bar{k})\}}^{(j)} < \frac{\pi}{2}.$$

We further define

$$R_j = \min_{\mathbf{h}_k^{(j)} \in U_j} \{R_{k,span\{U_j(\bar{k})\}}^{(j)}\}, \quad 1 \leq j \leq N_U \tag{6}$$

for each $U_j$, which essentially determines the system performance at high SNR [10][17].

In this paper, we mainly adopt the antenna selection rule that maximizes the post-processing SNR of the worst data stream (as in [1]). That is, we choose the subset among (4) such that $R_j$ in (6) is maximized, and we denote

$$R_{SL} = \max_{1 \leq j \leq N_U} \{R_j\}. \tag{7}$$

We will show that this selection rule is optimal for linear receivers with respect to diversity order after introducing some notation.

The diversity order of a communication system is defined as the slope of its error probability $P_e(\rho_0)$ in log-scale in the high SNR regime [7]:

$$d = -\lim_{\rho_0 \to \infty} \frac{\log P_e(\rho_0)}{\log(\rho_0)} = \lim_{\rho_0 \to \infty} \frac{\log P_e(\rho_0)}{\log(1/\rho_0)}. \tag{8}$$

Also, we adopt the operator $\doteq$ as defined in [7], to denote exponential equality, i.e. we write $f(\rho_0) \doteq \rho_0^{-b}$ to represent

$$-\lim_{\rho_0 \to \infty} \frac{\log f(\rho_0)}{\log(\rho_0)} = b.$$

Equivalently (for the convenience of analysis in this paper), we use $f(x) \doteq x^b$ to represent

$$\lim_{x \to 0} \frac{\log f(x)}{\log x} = b. \tag{9}$$

The operators $\dot{\leq}$, $\dot{\geq}$, $\dot{<}$, $\dot{>}$ are similarly defined. Note that according to our notation, $f(x) \dot{\leq} g(x)$ indicates $f(x) \geq g(x)$ for sufficiently small $x$.

**Lemma 1**: *For independently encoded spatial multiplexing systems with linear decorrelating receivers, the antenna selection method that chooses the antenna subset with the strongest weakest data link as in* (7), *achieves the optimal diversity order among all the antenna selection methods. Furthermore, the optimal diversity order can be evaluated as*

$$d_{opt}^{L} = \lim_{x \to 0} \frac{\log \Pr(R_{SL} \leq x)}{\log(x)}. \tag{10}$$

*Proof*: Suppose strategy $j$ is an arbitrary antenna selection rule, which may be channel dependent. Let the random variable $R_j$ denote the minimum squared projection height for this antenna selection rule. The conditional error probability of a layered SM system with a linear decorrelating receiver, after transmit antenna selection, can be upper and lower bounded as:

$$P_{e\_\max}(\mathbf{H}) \leq P_e(\mathbf{H}) \leq \sum_{l=1}^{L} P_{el}(\mathbf{H}), \tag{11}$$

where $P_{el}(\mathbf{H})$ is the error probability of the $l$th selected sub-stream, and $P_{e\_\max}(\mathbf{H}) = \max_l P_{el}(\mathbf{H})$ represents the worst of them (with instantaneous post-detection SNR of $\frac{\rho_0}{L} R_j(\mathbf{H})$). It is easily seen that

$$P_e \doteq P_{e\_\max} = E_{\mathbf{H}}\{P_{e\_\max}(\mathbf{H})\} = \int Q\left(\sqrt{\kappa \frac{\rho_0}{L} x}\right) f_{R_j}(x) dx, \tag{12}$$

where $Q(\cdot)$ denotes the Gaussian tail function, $\kappa$ is a modulation and coding dependent positive constant, and $f_{R_j}(\cdot)$ is the probability density function (PDF) of random variable $R_j$.

It is known [7][16] that the error probability is dominated by the outage probability, and the diversity order is given by

$$d = \lim_{x \to 0} \frac{\log \Pr(R_j \leq x)}{\log(x)}. \tag{13}$$

Our antenna selection rule (7) dictates $R_{SL} \geq R_j$ for any antenna selection rule $j$ with probability 1, so the lemma follows.

∎

Remark: This lemma applies to linear MMSE receivers as well. With appropriate modifications to the definitions of the link quantities, it can be extended to other linear receivers as well. However, as we will show below, this antenna selection rule is not optimal for decision-feedback receivers, though the optimal diversity orders are the same for both. We also make no claims on the practicality of this algorithm, as the main focus of this paper is on theoretical analysis. Note that efficient antenna selection algorithms exist in literature (e.g., [2][15] and references therein).

## III. Diversity Order and Diversity-Multiplexing Tradeoff when $L=2$

In this section, we discuss the main idea of our geometrical approach through the $L=2$ case for ease of illustration, which also admits an exact result:

**Theorem 1**: *In an $N_R \times N_T$ layered spatial multiplexing system with linear decorrelating/MMSE or decorrelating/MMSE decision-feedback receivers satisfying $N_T \geq 2$ and $N_R \geq 2$, if $L=2$ independently encoded data streams are transmitted from two selected antennas, the optimal achievable diversity order is $(N_T-1)(N_R-1)$.*

We start our proof with linear decorrelating receivers and then extend to decision-feedback receivers. These results can be readily applied to their MMSE counterparts as diversity order is a measure at high SNR. Finally we analyze the corresponding diversity-multiplexing tradeoff curves.

### A. Linear Receiver

For $L=2$, supposing that antennas $k$ and $j$ are selected, we have the following expression for the post-processing SNR corresponding to the data stream transmitted from antenna $k$ (c.f. (5)):

$$\rho_k = \frac{\rho_0}{L} R_{kj} = \frac{\rho_0}{L} \|\mathbf{h}_k\|^2 \sin^2 \theta_{kj}, \tag{14}$$

where we have abused the notations a bit without incurring ambiguity. It can be shown [18][19][12] that $\|\mathbf{h}_k\|^2$ is $\chi^2(2N_R)$ distributed, and $\theta_{kj}$ assumes a PDF of

$$f_{\theta_{kj}}(\theta) = (N_R-1)\sin 2\theta (\sin\theta)^{2N_R-4}, \quad \theta \in (0, \frac{\pi}{2}). \tag{15}$$

Furthermore, $\|\mathbf{h}_k\|^2$, $\|\mathbf{h}_j\|^2$ and $\theta_{kj}$ are mutually independent. Also, $R_{kj}$ is a $\chi^2(2(N_R-1))$ distributed random variable, i.e.

$$\Pr(R_{kj} \leq x) \doteq x^{N_R-1}. \tag{16}$$

Therefore the diversity order without antenna selection (or with random antenna selection) is $N_R-1$, while through optimal transmit antenna selection

$$R_{SL} = \max_{k \neq j \in \{1,\cdots,N_T\}} \{\min\{R_{kj}, R_{jk}\}\}, \tag{17}$$

we will show that a product gain of $N_T-1$ can be achieved.

In the following, the optimal diversity order (10) for linear receivers $d_{opt}^L$ will be explicitly explored. Note that neither the exact PDF of $R_{SL}$ nor its polynomial expansion near zero seems tractable, which motivates us to solve the problem through tight upper and lower bounds.

By definition,
$$\Pr(R_{SL} \leq x) = \Pr(\min(R_{12}, R_{21}) \leq x, \min(R_{13}, R_{31}) \leq x, \cdots, \min(R_{(N_T-1)N_T}, R_{N_T(N_T-1)}) \leq x)$$
$$= \Pr(\bigcup_{i=1}^{N} A_i), \quad (18)$$

where $N = 2^{N_U}$ events $\{A_i\}$ are defined as:
$$A_1 = \{R_{12} \leq x, R_{13} \leq x, \cdots, R_{1N_T} \leq x, R_{23} \leq x, \cdots, R_{(N_T-1)N_T} \leq x\}$$
$$A_2 = \{R_{12} \leq x, R_{13} \leq x, \cdots, R_{1N_T} \leq x, R_{23} \leq x, \cdots, R_{N_T(N_T-1)} \leq x\}$$
$$\vdots \quad (19)$$
$$A_N = \{R_{21} \leq x, R_{31} \leq x, \cdots, R_{N_T 1} \leq x, R_{32} \leq x, \cdots, R_{N_T(N_T-1)} \leq x\}.$$

Intuitively, the selection rule of (17) dictates that at least one element from each of the $N_U = \binom{N_T}{2}$ possible subsets should be in outage. Clearly we have:
$$\max \Pr(A_i) \leq \Pr(R_{SL} \leq x) \leq \sum_{i=1}^{N} \Pr(A_i), \quad (20)$$

which indicates
$$d_{opt}^L = \lim_{x \to 0} \frac{\log \max \Pr(A_i)}{\log(x)}. \quad (21)$$

However, it is generally difficult to identify the dominant terms at high SNR from (19), since the cardinality grows exponentially with $N_T^2$. Alternatively, we take the following approach. First, we find a common upper bound for $\Pr(A_i)$, which determines a lower bound for $d_{opt}^L$. We then evaluate the error exponential of $\Pr(A_1)$ (more precisely one of its lower bounds) and give an upper bound for $d_{opt}^L$. It turns out that these two bounds coincide and represent the best achievable diversity order.

The following Lemma is useful for obtaining an explicit lower bound of the diversity order.

**Lemma II**: *For any permutation of $1 \sim N_T$, denoted as $k_1 \sim k_{N_T}$, we have*
$$\Pr(R_{k_1 k_2} \leq x, R_{k_2 k_3} \leq x, \cdots, R_{k_{(N_T-1)} k_{N_T}} \leq x) = [\Pr(R_{kj} \leq x)]^{(N_T-1)}, \quad \forall k \neq j. \quad (22)$$

*Proof*: We need to show that random variables in the sequence $R_{k_1 k_2}, R_{k_2 k_3}, \cdots, R_{k_{(N_T-1)} k_{N_T}}$ are jointly independent. Essentially $R_{k_i k_{i+1}}$ is only a function of $\mathbf{h}_{k_i}$ and $\mathbf{h}_{k_{i+1}}$, denoted as $R_{k_i k_{i+1}} = g(\mathbf{h}_{k_i}, \mathbf{h}_{k_{i+1}})$, therefore the conditional PDF of $R_{k_i k_{i+1}}$ given those variables appearing earlier in the sequence admits:

$$f(R_{k_i k_{i+1}} | R_{k_{i-1} k_i}, \cdots, R_{k_1 k_2}) = f(g(\mathbf{h}_{k_i}, \mathbf{h}_{k_{i+1}}) | g(\mathbf{h}_{k_{i-1}}, \mathbf{h}_{k_i}), \cdots, g(\mathbf{h}_{k_1}, \mathbf{h}_{k_2}))$$
$$= f(g(\mathbf{h}_{k_i}, \mathbf{h}_{k_{i+1}}) | g(\mathbf{h}_{k_{i-1}}, \mathbf{h}_{k_i})) = f(R_{k_i k_{i+1}} | R_{k_{i-1} k_i}),$$

where the second equality holds because the states of $\mathbf{h}_{k_1} \sim \mathbf{h}_{k_{i-1}}$ do not affect $\mathbf{h}_{k_i}$ and $\mathbf{h}_{k_{i+1}}$. Therefore, the above sequence forms a Markov chain.

We are left to prove the independence between $R_{k_i k_{i+1}}$ and $R_{k_{i-1} k_i}$. Given $R_{k_i k_{i+1}} = \|\mathbf{h}_{k_i}\|^2 \sin^2 \theta_{k_{i+1} k_i}$ and $R_{k_{i-1} k_i} = \|\mathbf{h}_{k_{i-1}}\|^2 \sin^2 \theta_{k_{i-1} k_i}$, because of the independence between $\|\mathbf{h}_{k_i}\|^2$ and $\|\mathbf{h}_{k_{i-1}}\|^2$, and between vector norms and directions (angles) [18][19], we only need to show that $\theta_{k_{i+1} k_i}$ and $\theta_{k_{i-1} k_i}$ are independent.

Following a similar rotation approach as in [12], we define $\mathbf{e}_1 \sim \mathbf{e}_{N_R}$ as a *fixed* orthonormal basis (e.g., Cartesian coordinates) of the vector space $\mathbb{C}^{N_R}$. We rotate $[\mathbf{h}_{k_{i-1}}, \mathbf{h}_{k_i}, \mathbf{h}_{k_{i+1}}]$ as a whole so that $\mathbf{h}_{k_i}$ is parallel to $\mathbf{e}_1$, denoted as $[\tilde{\mathbf{h}}_{k_{i-1}}, \tilde{\mathbf{h}}_{k_{i+1}}] = \mathbf{Q}(\psi_{k_i})[\mathbf{h}_{k_{i-1}}, \mathbf{h}_{k_{i+1}}]$, where $\psi_{k_i}$ is the angle between $\mathbf{h}_{k_i}$ and $\mathbf{e}_1$, and $\mathbf{Q}(\psi_{k_i})$ is the corresponding unitary rotation matrix. Since $[\mathbf{h}_{k_{i-1}}, \mathbf{h}_{k_{i+1}}]$ is an i.i.d. Gaussian matrix (therefore the joint distribution is rotationally invariant) and is independent with $\psi_{k_i}$, $[\tilde{\mathbf{h}}_{k_{i-1}}, \tilde{\mathbf{h}}_{k_{i+1}}]$ is still i.i.d. Gaussian. Because $\theta_{k_{i+1} k_i}$ and $\theta_{k_{i-1} k_i}$ are unchanged after the rotation, and equal to the angles between $\tilde{\mathbf{h}}_{k_{i+1}}$ and $\mathbf{e}_1$, and between $\tilde{\mathbf{h}}_{k_{i-1}}$ and $\mathbf{e}_1$, respectively (see Figure 1), given the fact that $\tilde{\mathbf{h}}_{k_{i+1}}$ and $\tilde{\mathbf{h}}_{k_{i-1}}$ are independent, it is straightforward to show that $\theta_{k_{i+1} k_i}$ and $\theta_{k_{i-1} k_i}$ are independent, so are $R_{k_i k_{i+1}}$ and $R_{k_{i-1} k_i}$, and Lemma II follows.

∎

***Corollary I***: $\theta_{12}, \theta_{13}, \cdots, \theta_{1 N_T}$ *are jointly independent.*

*Proof*: Using the same rotation approach as above. If we rotate $[\mathbf{h}_1, \mathbf{h}_2, \cdots, \mathbf{h}_{N_T}]$ as a whole such that $\mathbf{h}_1$ is parallel to $\mathbf{e}_1$, $\tilde{\mathbf{h}}_2, \cdots, \tilde{\mathbf{h}}_{N_T}$ are jointly independent vectors, whose angles with $\mathbf{e}_1$, which equal to $\theta_{12}, \theta_{13}, \cdots, \theta_{1 N_T}$ respectively, are also jointly independent.

∎

Given Lemma II, a lower bound for the optimal diversity order (or an upper bound of the error performance at high SNR) is in order.

***Proposition I***: *The diversity order defined in (10) is lower bounded as* $d_{opt}^L \geq (N_T - 1)(N_R - 1)$.

*Proof:* Define $S_i$ as the set consisting of the $N_U = \binom{N_T}{2}$ random variables in $A_i$ (see (19)). For example, $S_1 = \{R_{12}, R_{13}, \cdots, R_{1N_T}, R_{23}, \cdots, R_{2N_T}, R_{34}, \cdots, R_{(N_T-1)N_T}\}$. A key observation is that in any $S_i$ we can always find an independent subset of $N_T - 1$ random variables bearing the same form as in (22). For example, in $S_1$, such a subset is given by $S_{1\_indep} = \{R_{12}, R_{23}, \cdots, R_{(N_T-1)N_T}\}$. By Lemma II, we can get the following upper bound for $\Pr(A_i)$:

$$\Pr(A_i) \leq [\Pr(R_{kj} \leq x)]^{(N_T-1)}, \quad \forall k \neq j, \quad \forall A_i. \tag{23}$$

With (16) we have

$$\max_i \Pr(A_i) \dot{\geq} x^{(N_T-1)(N_R-1)}, \tag{24}$$

and Proposition I follows.

∎

To find an upper bound for the optimal diversity order we choose to evaluate $\Pr(A_1)$. Intuitively $A_1$ contains the most correlated terms (projection heights from the same column vectors) and could potentially be one of the critical terms. The following result is obtained after some technically involved calculations, which verifies our conjecture.

**Proposition II**: *The diversity order defined in (10) is upper bounded as $d_{opt}^L \leq (N_T - 1)(N_R - 1)$.*

*Proof:* $\Pr(A_1)$ in (20) can be expressed as

$$\Pr(A_1) = \Pr(R_{12} \leq x, R_{13} \leq x, \cdots, R_{1N_T} \leq x, R_{23} \leq x, \cdots, R_{(N_T-1)N_T} \leq x)$$
$$= \Pr(\|\mathbf{h}_1\|^2 \sin^2 \theta_{12} \leq x, \cdots, \|\mathbf{h}_1\|^2 \sin^2 \theta_{1N_T} \leq x, \|\mathbf{h}_2\|^2 \sin^2 \theta_{23} \leq x, \cdots, \|\mathbf{h}_{N_T-1}\|^2 \sin^2 \theta_{(N_T-1)N_T} \leq x).$$

Defining $z = \sum_{k=1}^{N_T-1} \|\mathbf{h}_k\|^2$, which is distributed as $\chi^2(2(N_T-1)N_R)$, and $\psi_0 = (\pi/2)/(N_T-1)$, we have:

$$\Pr(A_1) \geq \Pr(z \sin^2 \theta_{12} \leq x, \cdots, z \sin^2 \theta_{1N_T} \leq x, z \sin^2 \theta_{23} \leq x, \cdots, z \sin^2 \theta_{(N_T-1)N_T} \leq x)$$
$$\geq \Pr(z \sin^2 \theta_{12} \leq x, \cdots, z \sin^2 \theta_{(N_T-1)N_T} \leq x, 0 < \theta_{12} < \psi_0, 0 < \theta_{13} < \psi_0, \cdots, 0 < \theta_{1N_T} < \psi_0), \tag{25}$$

where for the second inequality we have further restricted the ranges of the i.i.d. random variables $\theta_{12} \sim \theta_{1N_T}$ within $(0, \psi_0)$ (c.f. Corollary I). Based on the geometric structure involved, given $\theta_{12}$ and $\theta_{13}$, $\theta_{23}$ is constrained as $\theta_{23} \leq \theta_{12} + \theta_{13}$, where the equality holds only when $\mathbf{h}_1 \sim \mathbf{h}_3$ are linearly dependent (located in the same subspace with a dimension less than 3) [11]. Then within the range $(0, \psi_0)$, we have

$$\sin^2 \theta_{23} \leq \sin^2(\theta_{12} + \theta_{13}) \leq \sin^2(\theta_{12} + \theta_{13} + \cdots + \theta_{1N_T}) = \sin^2 \theta_\Sigma, \tag{26}$$

where $\theta_\Sigma = \sum_{k=2}^{N_T} \theta_{1k}$ is still in the range of $(0, \pi/2)$. Similar results hold for $\sin^2 \theta_{24} \sim \sin^2 \theta_{(N_T-1)N_T}$.

Therefore (25) can be further lower bounded as:

$$\begin{aligned} \Pr(A_1) &\geq \Pr(z \sin^2 \theta_\Sigma \leq x, 0 < \theta_{12} < \psi_0, 0 < \theta_{13} < \psi_0, \cdots, 0 < \theta_{1N_T} < \psi_0) \\ &\doteq \Pr(z \sin^2 \theta'_\Sigma \leq x), \end{aligned} \tag{27}$$

where we define a new set of i.i.d. random variables $\theta'_{12} \sim \theta'_{1N_T}$ with PDF of

$$f_{\theta'_{1i}}(x) = \frac{f_{\theta_{1i}}(x)}{\int_0^{\psi_0} f_{\theta_{1i}}(x) dx} = \frac{f_{\theta_{1i}}(x)}{C}, \quad 0 < x < \psi_0, \quad 2 \leq i \leq N_T, \tag{28}$$

i.e., the restriction of $\theta_{12} \sim \theta_{1N_T}$ in the range of $(0, \psi_0)$, and $\theta'_\Sigma = \sum_{i=2}^{N_T} \theta'_{1i}$.

Direct evaluation of (27) still seems intractable due to the involved PDF expression of $\theta'_\Sigma$. Alternatively, we further simplify it with some lemmas on exponential equivalence given in Appendix A. Specifically, with $\theta'_0 = \max_k \theta'_{1k}$ and $m(x) = \sin^2(x)$ for $x \in (0, \psi_0)$, by Lemma IV (whose proof requires Lemma III) in Appendix A, we have

$$\Pr(\sin^2 \theta'_\Sigma \leq x) \doteq \Pr(\sin^2 \theta'_0 \leq x). \tag{29}$$

Further by Lemma V in Appendix A and Appendix B, we have

$$\Pr(z \sin^2 \theta'_\Sigma \leq x) \doteq \Pr(z \sin^2 \theta'_0 \leq x) \doteq \Pr(z \sin^2 \theta_0 \leq x), \tag{30}$$

where $\theta_0 = \max_k \theta_{1k}$. We are left to evaluate the smallest exponential in $\Pr(z \sin^2 \theta_0 \leq x)$. Note that $z$ is a $\chi^2(2(N_T-1)N_R)$ distributed random variable with cumulative distribution function (CDF):

$$F_z(x) = 1 - e^{-x} \sum_{k=0}^{M+N_T-2} \frac{x^k}{k!}, \quad M = (N_T-1)(N_R-1), \tag{31}$$

while the PDF of $\theta_0$ is given by

$$f_{\theta_0}(\theta) = M[\sin^2 \theta]^{M-1} \sin 2\theta, \quad M = (N_T-1)(N_R-1), \quad \theta \in (0, \frac{\pi}{2}). \tag{32}$$

After some algebra presented in Appendix C, we can get the following equivalent polynomial form as $x \to 0$:

$$\Pr(z\sin^2\theta_0 \leq x) = \left(\frac{1}{M!} - M\sum_{k=0}^{N_T-3}\frac{k!}{(M+k+1)!}\right)x^M + o(x^M), \quad M = (N_T-1)(N_R-1), \tag{33}$$

where the coefficient of $x^M$ in (33) is always positive, which completes the proof.

∎

With Proposition I and II, Theorem I is proved for the linear receiver case. For example, when $N_T = 3$, $N_R = 3$, from (33) we get as $x \to 0$

$$\Pr(z\sin^2\theta_0 \leq x) = \frac{x^4}{120} + o(x^5).$$

Figure 2 presents the relevant quantities for $L=2$ and a linear decorrelating receiver, verifying a diversity order of 4.

## B. Decision-Feedback Receivers

Our geometrical analysis can also be applied for SM systems with decision-feedback receivers, whose performance is dominated by the first decoded layer, which is equivalent to a linear decorrelating (MMSE) receiver. For DF receivers with $L=2$, the system error probability is given by

$$P_e = P_{e1} + P_{e2}(1-P_{e1}), \tag{34}$$

where $P_{e1}$ is the error probability of the first decoded layer, and $P_{e2}$ is that of the second layer assuming perfect feedback. Therefore $P_e \doteq \max\{P_{e1}, P_{e2}\}$. For a fixed-order DF receiver without antenna selection, $P_{e1} \geq P_{e2}$ is always fulfilled, so that we can investigate its diversity order solely from the first layer (see, e.g. [12][17]). *However, this situation is not always true in the antenna selection context.* In this subsection we will derive the optimal achievable diversity order for independently encoded SM systems with DF receivers and transmit antenna selection in the following way. First we will show that the optimal achievable diversity order of the first layer $d_{1,opt}^{DF} = (N_T-1)(N_R-1)$, following a procedure similar to what we have discussed above for linear receivers; therefore $d_{opt}^{DF} \leq d_{1,opt}^{DF} = (N_T-1)(N_R-1)$. But this antenna selection rule, while optimal for linear receivers, is *not optimal* for decision-feedback receivers. Therefore, we continue by constructing a specific antenna selection algorithm which can achieve a diversity order of $(N_T-1)(N_R-1)$, so that $d_{opt}^{DF} \geq (N_T-1)(N_R-1)$.

### B1. Transmit Antenna Selection that Maximizes the SNR of the First Layer

We investigate an antenna selection algorithm similar to the one in III.A: selecting the antenna subset that maximizes the post-processing SNR of the first decoded layer. We distinguish two scenarios with respect to detection order: arbitrary but fixed ordering and optimal ordering [9]. For the former case, without loss of generality, we assume that the decoding starts from the signal transmitted from the antenna with the smallest index number in the selected antenna subset. The first layer diversity order can be expressed as

$$d_1 = \lim_{\rho_0 \to \infty} \frac{P_{e1}(\rho_0)}{\log(1/\rho_0)} = \lim_{\rho_0 \to \infty} \frac{P_{out1}(\rho_0)}{\log(1/\rho_0)} = \lim_{x \to 0} \frac{\Pr(R_{SL1} \le x)}{\log(x)},$$

where $P_{out1}$ is the outage probability of the first layer, whose post-processing SNR is proportional to $R_{SL1}$, defined as (c.f. (17))

$$R_{SL1} = \max_{k<j \in \{1,\cdots,N_T\}} \{R_{kj}\}. \tag{35}$$

Clearly we have (c.f. (19))

$$\Pr(R_{SL1} \le x) = \Pr(R_{12} \le x, R_{13} \le x, \cdots, R_{1N_T} \le x, R_{23} \le x, \cdots, R_{(N_T-1)N_T} \le x) = \Pr(A_1), \tag{36}$$

whose upper bound can be derived from (23), while its lower bound exponential behavior evaluation directly follows (25)~(33). Therefore for arbitrary but fixed ordering

$$d_{1,opt}^{DF} = (N_T - 1)(N_R - 1). \tag{37}$$

For the optimal ordering case, our selection rule is reformulated with

$$R_{SL1} = \max_{k \ne j \in \{1,\cdots,N_T\}} \{\max\{R_{kj}, R_{jk}\}\}, \tag{38}$$

and we have:

$$\begin{aligned}\Pr(R_{SL1} \le x) &= \Pr(\max(R_{12}, R_{21}) \le x, \max(R_{13}, R_{31}) \le x, \cdots, \max(R_{(N_T-1)N_T}, R_{N_T(N_T-1)}) \le x) \\ &= \Pr(\max(\|\mathbf{h}_1\|^2, \|\mathbf{h}_2\|^2)\sin^2\theta_{12} \le x, \cdots, \max(\|\mathbf{h}_{N_T-1}\|^2, \|\mathbf{h}_{N_T}\|^2)\sin^2\theta_{(N_T-1)N_T} \le x).\end{aligned} \tag{39}$$

Following the same geometrical approach, it is straightforward to upper and lower bound (39) as:

$$\Pr(R_{SL1} \le x) \le \Pr(A_1) \le \left[\Pr(R_{kj} \le x)\right]^{N_T - 1}, \quad \forall k \ne j, \tag{40}$$

and

$$\begin{aligned}\Pr(R_{SL1} \le x) &\ge \Pr(\max(\|\mathbf{h}_1\|^2, \cdots, \|\mathbf{h}_{N_T}\|^2)\sin^2\theta'_{\Sigma} \le x) \\ &\ge \Pr((\sum_{i=1}^{N_T}\|\mathbf{h}_i\|^2)\sin^2\theta'_{\Sigma} \le x).\end{aligned} \tag{41}$$

The evaluation of the lower bound in (41) is similar as in (27), except that $z$ is re-defined as $z = \sum_{k=1}^{N_T}\|\mathbf{h}_k\|^2$. Following a similar approach as (27)~(33), we get

$$\Pr(R_{SL1} \le x) \le \left(\frac{1}{M!} - M \sum_{k=0}^{N_T+N_R-3} \frac{k!}{(M+k+1)!}\right)x^M + o(x^{M+1}), \quad M = (N_T-1)(N_R-1). \tag{42}$$

So the same result as in (37) is obtained with optimal ordering. In [12], the authors have shown that the optimal ordering will not increase the diversity in the first layer of a SM system with DF receivers. As a side product, here we present the same result in the antenna selection context. To summarize we have the following proposition.

***Proposition III***: *The optimal diversity order of the first decoded layer of an independently coded SM system with DF receivers is $(N_T-1)(N_R-1)$, either with or without optimal ordering. As a result, the optimal diversity order of such a system is upper bounded as $d_{opt}^{DF} \leq (N_T-1)(N_R-1)$.*

Remark: Although this antenna selection algorithm guarantees the best error performance for the first layer, it is in general not optimal with respect to the diversity order for the whole system. The reason is that although $P_{e1}$ in (34) is maximized, $P_{e2}$ is not affected by the selection process. Rather, it behaves the same as in a non-selection scheme with $N_R$-order diversity. Therefore $P_e \doteq \max\{P_{e1}, P_{e2}\}$ is mostly dominated by the second layer, and the diversity order is given by $d = \min\{(N_T-1)(N_R-1), N_R\} = N_R$, for $N_T \geq 3$, $N_R \geq 2$.

In the next subsection, we analyze a simple yet effective antenna selection algorithm, for which the first decoded layer achieves a diversity order of $(N_T-1)(N_R-1)$, while the second layer performs better than the first layer.

**B2. A QR Decomposition Based Antenna Selection Algorithm**

This antenna selection algorithm is based on QR decomposition, which was originally proposed for capacity maximization [14][15]. Compared with brute force methods, this algorithm greatly reduces the computational complexity while achieving a near optimal performance with respect to channel capacity. Here we apply it in SM systems with DF receivers with the goal of minimizing the error rate, and show that it is also optimal with respect to diversity order, therefore verifying our observations in [15] and revealing its great potential.

From a geometrical viewpoint, this incremental antenna selection procedure starts by seeking the column vector with the largest norm (or the largest projection height to the null-space); and in each of the following steps, one column with the largest projection height to the space spanned by the selected column vectors is chosen until the *L*th antenna is selected. The detection order is the reverse of the selection order, i.e., the stream decoded at the *l*th step is transmitted from the antenna selected at the $L-l+1$ step. The corresponding post-processing SNR for the stream decoded at the *l*th step is then proportional to the maximum value among $N_T - L + l$ independent random variables distributed as $\chi^2(2(N_R - L + l))$, resulting in a diversity order of $(N_T - L + l)(N_R - L + l)$. Therefore for $L = 2$, the first decoded layer achieves a diversity order of $d_1 = (N_T - 1)(N_R - 1)$ and the second decoded layer achieves $d_2 = N_T N_R$, and the diversity order for the whole system equals to $(N_T - 1)(N_R - 1)$. Therefore we have the following result.

***Proposition IV***: The optimal diversity order of an independently coded SM system with DF receivers is lower bounded by $d_{opt}^{DF} \geq (N_T - 1)(N_R - 1)$.

Remark: This QR based antenna selection algorithm is optimal for independently encoded SM systems with DF receivers with respect to diversity order. However, it is in general not optimal for linear receivers. This interesting fact should be contrasted with remarks for Lemma I and Proposition III. From the simulation result in Figure 3 for a $N_T = 3$, $N_R = 3$, $L = 2$ scenario, we see that although simpler, the QR based method outperforms the algorithm that optimizes only the first layer (bearing a higher diversity).

With Proposition III and IV, Theorem I is also proved for DF receivers.

## C. The Diversity-Multiplexing Tradeoff

Using the same geometrical approach, we can also obtain the diversity-multiplexing tradeoff curve introduced in [7] for the independently encoded SM systems with antenna selection. With quasi-static fading assumption, a family of codes $\{\varsigma(\rho_0)\}$ over a block length shorter than fading coherence time is employed, one at each SNR level. We further assume that the rate of the code increases with SNR, so a scheme achieves a multiplexing gain $r$ if the rate $R(\rho_0) = r \log \rho_0$. Based on the diversity order analysis in Section III.A and B, assuming equal-power and equal rate allocation, the following result is in order.

***Theorem II:*** Under the same setting as Theorem I, the optimal diversity-multiplexing tradeoff curve is the piecewise function of r connecting the points $(r, d_{opt}(r))$ for $0 \leq r \leq 2$, with

$$d_{opt}(r) = (N_T - 1)(N_R - 1)(1 - r/2)^+, \tag{43}$$

where $(x)^+ = \max(x, 0)$.

*Proof*: Based on [7] and our previous analysis

$$P_e \doteq P_{e\_max} \doteq P_{out\_max},$$

where $P_{out\_max}$ can be viewed as the outage probability of the worst stream for linear receivers, and that of the first decoded layer for DF receivers, respectively. Therefore the diversity-multiplexing tradeoff curve $d(r)$ can be evaluated directly from $P_{out\_max}$ as $\rho_0 \to \infty$:

$$P_{out\_max} = \Pr\left[\log(1+\frac{\rho_0}{L}R_{SL}) \le \frac{r}{L}\log\rho_0\right] \simeq \Pr\left[R_{SL} \le L\rho_0^{-(1-\frac{r}{L})}\right]$$

$$\doteq \left[\rho_0^{-(1-\frac{r}{L})}\right]^M = \rho_0^{-M(1-\frac{r}{L})},$$

where $M = (N_T - 1)(N_R - 1)$.

■

## IV. Extension to General $L$ Scenarios

The analysis for general $L$ follows a similar approach as the $L=2$ case in Section III. However, the evaluation becomes more involved as now the post-processing SNR is proportional to the squared projection height from a column vector to a non-degenerated space. We can only obtain upper and lower bounds for the achievable diversity order as follows.

***Theorem III:*** *In an $N_R \times N_T$ layered spatial multiplexing system with linear decorrelating/MMSE or decorrelating/MMSE decision-feedback receivers satisfying $N_T \ge L$ and $N_R \ge L$, if independently encoded data streams are transmitted from $L$ selected antennas, the optimal achievable diversity order is bounded as*

$$M_L \le d \le M_U, \tag{44}$$

*where $M_L = (N_T - L + 1)(N_R - L + 1)$, and $M_U = (N_T - L + 1)(N_R - 1)$; while the optimal diversity-multiplexing tradeoff curve is bounded as:*

$$M_L(1-\frac{r}{L})^+ \le d(r) \le M_U(1-\frac{r}{L})^+. \tag{45}$$

Again we start with linear receivers. Employing the same antenna selection method as introduced for $L=2$ scenarios (maximizing the weakest link), we can then derive a similar outage probability expression as (18):

$$\Pr(R_{SL} \le x) = \Pr(\min_{\mathbf{h}_k^{(1)} \in U_1}\{R^{(1)}_{k,span\{U_1(\bar{k})\}}\} \le x, \min_{\mathbf{h}_k^{(2)} \in U_2}\{R^{(2)}_{k,span\{U_2(\bar{k})\}}\} \le x, \cdots, \min_{\mathbf{h}_k^{(N_U)} \in U_{N_U}}\{R^{(N_U)}_{k,span\{U_{N_U}(\bar{k})\}}\} \le x), \tag{46}$$

where $U_1 \sim U_{N_U}$ are defined in (4). By defining the following events with $N = L\binom{N_T}{L}$ (c.f. (19)):

$$A_1 = \{R^{(1)}_{1,span\{U_1(\bar{1})\}} \le x, \cdots, R^{(N_1)}_{1,span\{U_{N_1}(\bar{1})\}} \le x, R^{(N_1+1)}_{1,span\{U_{N_1+1}(\bar{1})\}} \le x, \cdots, R^{(N_U)}_{1,span\{U_{N_U}(\bar{1})\}} \le x\}$$

$$A_2 = \{R^{(1)}_{1,span\{U_1(\bar{1})\}} \le x, \cdots, R^{(N_1)}_{1,span\{U_{N_1}(\bar{1})\}} \le x, R^{(N_1+1)}_{1,span\{U_{N_1+1}(\bar{1})\}} \le x, \cdots, R^{(N_U)}_{2,span\{U_{N_U}(\bar{2})\}} \le x\}$$

$$\vdots$$

$$A_N = \{R^{(1)}_{L,span\{U_1(\bar{L})\}} \leq x, \cdots, R^{(N_1)}_{L,span\{U_{N_1}(\bar{L})\}} \leq x, R^{(N_1+1)}_{L,span\{U_{N_1+1}(\bar{L})\}} \leq x, \cdots, R^{(N_U)}_{L,span\{U_{N_U}(\bar{L})\}} \leq x\},$$

(20) still holds.

We have the following extension of Lemma II for general $L^3$, where we have abused the notations a bit, using $span\{k_2, k_3, \cdots, k_L\}$ to denote $span\{\mathbf{h}_{k_2}, \mathbf{h}_{k_3}, \cdots, \mathbf{h}_{k_L}\}$.

**Lemma VI**: *For any permutation of $1 \sim N_T$, denoted as $k_1 \sim k_{N_T}$, we have:*

$$\Pr(R_{k_1,span\{k_2,k_3,\cdots,k_L\}} \leq x, R_{k_2,span\{k_3,k_4,\cdots,k_{L+1}\}} \leq x, \cdots, R_{k_{N_T-L+1},span\{k_{N_T-L+2},k_{N_T-L+3},\cdots,k_{N_T}\}} \leq x\}$$
$$= [\Pr(R_{k_1,span\{k_2,k_3,\cdots,k_L\}} \leq x)]^{(N_T-L+1)}. \qquad (47)$$

*Proof*: From the definition, $R_{k_1,span\{k_2,k_3,\cdots,k_L\}} = \|\mathbf{P}\mathbf{h}_{k_1}\|^2$, where $\mathbf{P} = \mathbf{I} - \mathbf{B}\mathbf{B}^\dagger$ is the projection matrix to the null space of $span\{k_2, k_3, \cdots, k_L\}$, in which $\mathbf{B}$ is composed of any basis of $span\{k_2, k_3, \cdots, k_L\}$ [21]. Since any projection matrix is idempotent, i.e. $\mathbf{P}^2 = \mathbf{P}$, its eigenvalues are either 1 or 0. Noting that $\mathbf{P}$ is Hermitian, we can write the eigenvalue decomposition of $\mathbf{P}$ as $\mathbf{P} = \mathbf{V}\mathbf{\Lambda}\mathbf{V}^*$, where $\mathbf{V}$ is unitary, and $\mathbf{\Lambda} = diag(1^{N_R-L+1}, 0^{L-1})$. Therefore

$$R_{k_1,span\{k_2,k_3,\cdots,k_L\}} = \|\mathbf{V}\mathbf{\Lambda}\mathbf{V}^*\mathbf{h}_{k_1}\|^2 = \|\mathbf{\Lambda}\mathbf{V}^*\mathbf{h}_{k_1}\|^2, \qquad (48)$$

where the second equality follows by the fact that a unitary transformation preserves length. From the definition of $\mathbf{P}$, the unitary matrix $\mathbf{V}^*$ is independent of $\mathbf{h}_{k_1}$. Therefore

$$f(R_{k_1,span\{k_2,k_3,\cdots,k_L\}} | \mathbf{h}_{k_2}, \mathbf{h}_{k_3}, \cdots, \mathbf{h}_{k_L}) = f(\|\mathbf{\Lambda}\mathbf{V}_0^*\mathbf{h}_{k_1}\|^2) = f(\|\mathbf{\Lambda}\mathbf{h}_{k_1}\|^2) = f(R_{k_1,span\{k_2,k_3,\cdots,k_L\}}), \qquad (49)$$

where $\mathbf{V}_0^*$ is a fixed matrix dependent on the given realizations of $\mathbf{h}_{k_2}, \mathbf{h}_{k_3}, \cdots, \mathbf{h}_{k_L}$, and the second equality comes from the rotationally invariant property of the i.i.d. Gaussian vector $\mathbf{h}_{k_1}$ [19]. That is, $R_{k_1,span\{k_2,k_3,\cdots,k_L\}}$ is independent of $\mathbf{h}_{k_2}, \mathbf{h}_{k_3}, \cdots, \mathbf{h}_{k_L}$. It is then straightforward to show that $R_{k_1,span\{k_2,k_3,\cdots,k_L\}}$, $R_{k_2,span\{k_3,k_4,\cdots,k_{L+1}\}}$, $R_{k_3,span\{k_4,k_5,\cdots,k_{L+2}\}}$ ...... are jointly independent and Lemma VI holds. ∎

A lower bound of the optimal diversity order is then given by the following proposition:

**Proposition V**: *For general L, the optimal diversity order for linear receivers can be lower bounded as $d_{opt}^L \geq (N_T - L+1)(N_R - L+1)$.*

---

[3] Lemma II is a special case of Lemma VI. However, the proof of Lemma II bears some interesting geometric elements, and is needed for Corollary I and Proposition II.

*Proof*: Defining $S_i$ as the set collecting the $N_U = \binom{N_T}{L}$ random variables in $A_i$. We observe that in any $S_i$ we can always find a subset of $N_T - L + 1$ i.i.d. random variables bearing the same form as in (47). For example, when $N_T = 5$, $L = 3$, in $S_1$, such a subset is given by $S_{1\_indep} = \{R_{1,23}, R_{2,34}, R_{3,45}\}$. Therefore by Lemma VI, we can get the following upper bound for any $\Pr(A_i)$:

$$\Pr(A_i) \leq \left[\Pr(R^{(j)}_{k,span\{U_j(\bar{k})\}} \leq x)\right]^{N_T - L + 1}. \tag{50}$$

Furthermore, since $R^{(j)}_{k,span\{U_j(\bar{k})\}}$ is $\chi^2(2(N_R - L + 1))$ distributed, we have

$$\Pr(A_i) \dot{\geq} x^{(N_T - L + 1)(N_R - L + 1)}, \tag{51}$$

and Proposition V is proved.

∎

On the other hand, for $L > 2$ scenarios, the derivation of a tight lower bound for $\Pr(A_1)$ is much more involved as compared to the $L = 2$ case, because the angles $\{\theta^{(1)}_{k,span\{U_j(\bar{k})\}}\}$ are correlated in a complicated manner, and a general form of their joint PDF expressions is not accessible. Nonetheless, we have the following result.

***Proposition VI***: *For general L, the optimal diversity order for linear receivers can be upper bounded as* $d^L_{opt} \leq (N_T - L + 1)(N_R - 1)$.

*Proof*: Since the projection height from a vector to a subspace represents the shortest distance from the vector to any point in the subspace, we have $R^{(j)}_{k,span\{U_j(\bar{k})\}} \leq R^{(j)}_{kl}$, for any $\mathbf{h}^{(j)}_k$, $\mathbf{h}^{(j)}_l \in U_j$, where $R^{(j)}_{kl}$ denotes the squared projection height from $\mathbf{h}^{(j)}_k$ to $\mathbf{h}^{(j)}_l$. It is then not difficult to build up the following lower bound:

$$\begin{aligned}\Pr(A_1) &= \Pr(R^{(1)}_{1,span\{U_1(\bar{1})\}} \leq x, \cdots, R^{(N_1)}_{1,span\{U_{N_1}(\bar{1})\}} \leq x, R^{(N_1+1)}_{1,span\{U_{N_1+1}(\bar{1})\}} \leq x, \cdots, R^{(N_U)}_{1,span\{U_{N_U}(\bar{1})\}} \leq x) \\ &\geq \Pr(R^{(1)}_{12} \leq x, R^{(2)}_{12} \leq x, \cdots, R^{(N_U)}_{12} \leq x).\end{aligned} \tag{52}$$

Carefully examining the first two elements in all subsets (see (4)) reveals that the last expression in (52) bears a similar form as the $\Pr(A_1)$ in $L = 2$ case (see (19)), replacing $N_T - 1$ with $N_T - L + 1$. Following the same lines as in Section III, we have for general $L$

$$\Pr(R_{SL} \leq x) \leq \Pr(A_1) \dot{\leq} x^{(N_T - L + 1)(N_R - 1)}. \tag{53}$$

Combining the above two propositions, (44) is proved for linear receivers. Also, as for the $L=2$ case it is straightforward to show that (44) also holds for the first decoded layer of DF receivers. Since the diversity order of the first decoded layer is upper bounded by $(N_T - L + 1)(N_R - 1)$ and $P_e \leq P_{e1}$, the upper bound $(N_T - L + 1)(N_R - 1)$ also applies for $P_e$ with DF receivers. On the other hand, by employing the QR based selection algorithm for DF receivers, a diversity order lower bound $(N_T - L + 1)(N_R - L + 1)$ is achieved. Therefore (44) also holds for DF receivers. The derivation of (45) follows the same method as in Section III.C.

Note that when $L = 2$, the two bounds in (44) and (45) coincide and conform to the results obtained in Section III. We also conjecture that the lower bounds in (44) and (45) are tight and the key lies on finding a better lower bound of $\Pr(A_1)$ than (52).

A conjecture on the diversity order of independently encoded SM systems with transmit antenna selection and linear decorrelating receivers was made in [1] based on numerical results, which actually has motivated our research:

***Conjecture 1*** [1]: *For linear decorrelating receivers, when $N_R = L$, the achievable diversity order is $N_T - L + 1$.*

Our results prove its correctness and further extend it to general $N_R$ scenarios.

## V. Conclusions

In this paper, we have analyzed the diversity order achieved by transmit antenna selection for practical SM systems with linear and DF receivers. Using a geometric approach, we have rigorously derived their achievable diversity order for the $L = 2$ scenario. We also used the same geometrical approach to obtain bounds on the achievable diversity order for general *L*. Our results prove and extend the previous conjectures in literature drawn from simulations, and verify the predicted potential of antenna selection for practical spatial multiplexing systems.

The analysis for maximum-likelihood receivers, joint transmitter and receiver selection, and multiuser MIMO scenarios direct our future research.

# Appendix: Some Technical Results in the Proof of Proposition II

Appendix A: Some Lemmas on Exponential Equivalence

**Lemma III**: *If $\theta_1,\cdots,\theta_K$ are independent positive random variables, whose cumulative distribution functions admit $F_{\theta_k}(x) \doteq x^{n_k}$, we have*

$$\Pr\left[\sum_{k=1}^{K}\theta_k \leq x\right] \doteq x^{\sum_{k=1}^{K} n_k}. \tag{54}$$

*Proof*: At first we evaluate the CDF of $\theta_1 + \theta_2$ as $x \to 0$:

$$\begin{aligned}
\Pr(\theta_1 + \theta_2 \leq x) &= \int_0^x \left[\int_0^{x-\theta_1} f_{\theta_2}(\theta_2)d\theta_2\right] f_{\theta_1}(\theta_1)d\theta_1 \\
&= \int_0^x f_{\theta_1}(\theta_1) F_{\theta_2}(x-\theta_1) d\theta_1 \\
&\doteq \int_0^x \left[\theta_1^{n_1-1} + o(\theta_1^{n_1-1})\right]\cdot\left[(x-\theta_1)^{n_2} + o((x-\theta_1)^{n_2})\right]d\theta_1, \\
&\doteq x^{n_2} \int_0^x \left[\theta_1^{n_1-1} + o(\theta_1^{n_1-1})\right]d\theta_1 \\
&\doteq x^{n_1+n_2},
\end{aligned}$$

where $o(x)$ denotes a function of $x$ such that $\lim_{x\to 0} o(x)/x = 0$. Next, we can treat $\theta_1 + \theta_2$ as a new random variable exponentially equivalent to $x^{n_1+n_2}$ and evaluate $(\theta_1 + \theta_2) + \theta_2$ following the same approach, whose CDF asymptotically behaves as $x^{(n_1+n_2)+n_3}$. Lemma III follows after such repeated operations.

∎

**Lemma IV**: *Let $m(\theta)$ be a positive function of $\theta$ and monotonically increases with $\theta$, satisfying $m(0) = 0$ and $m^{-1}(x) \doteq x^{n_0}$. If $\theta_1,\cdots,\theta_K$ are independent positive random variables, whose cumulative distribution functions admit $F_{\theta_k}(x) \doteq x^{n_k}$, we have*

$$\Pr\left[m\left(\sum_{k=1}^{K}\theta_k\right) \leq x\right] \doteq \Pr\left[m\left(\max_k \theta_k\right) \leq x\right] \doteq x^{n_0 \sum_{k=1}^{K} n_k}. \tag{55}$$

*Proof*: Since $\theta_1,\cdots,\theta_K$ are independent, we have:

$$\Pr[\max_k \theta_k \leq x] = \prod_{k=1}^{K} F_{\theta_k}(x) \doteq x^{\sum_{k=1}^{K} n_k}. \qquad (56)$$

By Lemma III we have

$$\Pr[\sum_{k=1}^{K} \theta_k \leq x] \doteq x^{\sum_{k=1}^{K} n_k}. \qquad (57)$$

Because both $\theta$ and $m(\theta)$ are positive, and $m(\theta)$ monotonically increases with $\theta$, $m^{-1}(x)$ is also a positive function of $x$, which monotonically increases with $x$. Therefore,

$$\Pr[m(\max_k \theta_k) \leq x] = \Pr[\max_k \theta_k \leq m^{-1}(x)] \doteq [m^{-1}(x)]^{\sum_{k=1}^{K} n_k} \doteq x^{n_0 \sum_{k=1}^{K} n_k}. \qquad (58)$$

Similarly,

$$\Pr\left[m\left(\sum_{k=1}^{K} \theta_k\right) \leq x\right] \doteq x^{n_0 \sum_{k=1}^{K} n_k} \qquad (59)$$

and Lemma IV follows.

∎

**Lemma V:** For independent continuous random variables $a$, $b_1$ and $b_2$ satisfying $a \geq 0$ with $\Pr(a \leq x) \doteq x^{n_a}$, and $0 \leq b_1, b_2 \leq 1$ with $\Pr(b_1 \leq x) \doteq \Pr(b_2 \leq x) \doteq x^{n_b}$, we have

$$\Pr(ab_1 \leq x) \doteq \Pr(ab_2 \leq x) \leq x^{n_a}. \qquad (60)$$

*Proof:* We have

$$\begin{aligned}
\Pr(ab_1 \leq x) &= \int_0^1 \Pr(a \leq \frac{x}{y}) f_{b_1}(y).dy \\
&= \int_0^\varepsilon \Pr(a \leq \frac{x}{y}) f_{b_1}(y).dy + \int_\varepsilon^1 \Pr(a \leq \frac{x}{y}) f_{b_1}(y).dy \\
&= t_{11}(x) + t_{12}(x),
\end{aligned} \qquad (61)$$

and similarly,

$$\begin{aligned}
\Pr(ab_2 \leq x) &= \int_0^\varepsilon \Pr(a \leq \frac{x}{y}) f_{b_2}(y).dy + \int_\varepsilon^1 \Pr(a \leq \frac{x}{y}) f_{b_2}(y).dy \\
&= t_{21}(x) + t_{22}(x),
\end{aligned} \qquad (62)$$

where $\varepsilon > 0$ is a fixed small positive number, small enough to make the PDF approximations for $b_1$ and $b_2$:

$$f_{b_1}(x) = c_1 x^{n_b-1} + o(x^{n_b-1}), \quad f_{b_2}(x) = c_2 x^{n_b-1} + o(x^{n_b-1}) \tag{63}$$

hold true.

Since $\Pr(a \leq \frac{x}{y})$ is positive, and is a decreasing function of $y$, when $x \to 0$ we have

$$t_{12}(x) \leq \Pr(a \leq \frac{x}{\varepsilon}) \int_\varepsilon^1 f_{b_1}(y).dy \leq \Pr(a \leq \frac{x}{\varepsilon}) \doteq x^{n_a}. \tag{64}$$

In another word, $t_{12}(x) \dot\geq x^{n_a}$.

On the other hand, since $0 \leq b_1 \leq 1$, we have

$$\Pr(ab_1 \leq x) \geq \Pr(a \leq x) \doteq x^{n_a}, \tag{65}$$

which means that $\Pr(ab_1 \leq x) \dot\leq x^{n_a}$.

The same inequalities as in (64)(65) also hold true for $t_{22}(x)$ and $\Pr(ab_2 \leq x)$, respectively.

Meanwhile, since $\varepsilon$ is small enough, by applying (63), we get that

$$t_{11}(x) \doteq t_{21}(x) \doteq \int_0^\varepsilon \Pr(a \leq \frac{x}{y}) y^{n_b-1}.dy. \tag{66}$$

If $t_{12}(x) \dot> x^{n_a}$ or $t_{22}(x) \dot> x^{n_a}$, $t_{11}(x) \doteq t_{21}(x) \dot\leq x^{n_a}$ dominate and (60) follows. We are left to check the case $t_{12}(x) \doteq t_{22}(x) \doteq x^{n_a}$, which together with (66) leads to (60) as well.

∎

Appendix B: $\Pr(z \sin^2 \theta'_0 \leq x) \doteq \Pr(z \sin^2 \theta_0 \leq x)$

From (15)(28) the PDF of $\theta'_0 = \max_k \theta'_{1k}$ is derived through results in order statistics:

$$f_{\theta'_0}(\theta) = \frac{M}{C^{N_T-1}} [\sin^2 \theta]^{M-1} \sin 2\theta, \quad M = (N_T-1)(N_R-1), \quad \theta \in (0, \psi_0), \tag{67}$$

where $C = \int_0^{\psi_0} f_{\theta_{1i}}(x) dx, \ 2 \leq i \leq N_T$. Therefore, we have

$$\Pr(z \sin^2 \theta'_0 \leq x) = \int_0^{\psi_0} F_z\left(\frac{x}{\sin^2 \theta}\right) f_{\theta'_0}(\theta).d\theta$$

$$= \frac{M}{C^{N_T-1}} \int_0^{\psi_0} F_z\left(\frac{x}{\sin^2 \theta}\right) [\sin^2 \theta]^{M-1} \sin 2\theta.d\theta$$

$$\stackrel{t=\sin^2\theta}{=} \frac{M}{C^{N_T-1}} \int_0^{a_0} F_z\left(\frac{x}{t}\right) t^{M-1}.dt$$

$$\stackrel{t_1=\frac{t}{a_0}}{=} \frac{Ma_0^M}{C^{N_T-1}} \int_0^1 F_z\left(\frac{(x/a_0)}{t_1}\right) t_1^{M-1}.dt_1,$$

(68)

where $a_0 = \sin^2 \psi_0$ is a positive real number. On the other hand, by applying (32) we have

$$\Pr(z \sin^2 \theta_0 \leq x) = \int_0^{\pi/2} F_z\left(\frac{x}{\sin^2 \theta}\right) f_{\theta_0}(\theta).d\theta$$

$$= M \int_0^{\pi/2} F_z\left(\frac{x}{\sin^2 \theta}\right) [\sin^2 \theta]^{M-1} \sin 2\theta.d\theta$$

(69)

$$\stackrel{t=\sin^2\theta}{=} M \int_0^1 F_z\left(\frac{x}{t}\right) t^{M-1}.dt.$$

By comparing (68) and (69), we get $\Pr(z \sin^2 \theta'_0 \leq x) \doteq \Pr(z \sin^2 \theta_0 \leq x)$.

Appendix C: The Derivation of (33)

We continue the evaluation of (69) to derive the polynomial expansion of $\Pr(z \sin^2 \theta_0 \leq x)$:

$$\Pr(z \sin^2 \theta_0 \leq x) = M \int_0^1 F_z\left(\frac{x}{t}\right) t^{M-1}.dt$$

$$\stackrel{m=1/t}{=} M \int_1^\infty F_z(mx) \frac{1}{m^{M+1}}.dm$$

$$= M \int_1^\infty \left[1 - e^{-mx} \sum_{k=0}^{M+N_T-2} \frac{m^k x^k}{k!}\right] \frac{1}{m^{M+1}}.dm$$

$$= 1 - P_1(x) - P_2(x),$$

(70)

where $P_1(x) = M \sum_{k=0}^{M} \frac{x^k}{k!} E_{M+1-k}(x)$, and $P_2(x) = M \sum_{k=M+1}^{M+N_T-2} \frac{x^k}{k!} E_{-(k-M-1)}(x)$, with $E_k(x) = \int_1^\infty \frac{e^{-xm}}{m^k} dm$ the

integral exponential function. From [13], we have the following recursive rules:

$$E_{k+1}(x) = \frac{1}{k}(e^{-x} - xE_k(x))$$

$$E_{k+1}(x) = \frac{e^{-x}}{k!} \sum_{i=0}^{k-1} (-1)^i (k-i-1)! x^i + \frac{(-1)^k}{k!} x^k E_1(x).$$

Therefore, after some involved mathematical manipulations, we have

$$P_1(x) = Me^{-x} \sum_{k=0}^{M-1} c_k x^k,$$

where $c_k = \sum_{i=0}^{k} \dfrac{(-1)^{k-i}(M-k-1)!}{(M-i)!i!}$, for $k \leq M-1$. We can expand $e^{-x}$ by its Taylor's series and get the polynomial equivalent form: $P_1(x) = \sum_{n=0}^{\infty} a_n x^n$. For $n \leq M-1$,

$$\begin{aligned}
a_n &= M \sum_{k=0}^{n} \frac{(-1)^{n-k}}{(n-k)!} \sum_{i=0}^{k} \frac{(-1)^{k-i}(M-k-1)!}{(M-i)!i!} \\
&= M \frac{1}{n!} \sum_{k=0}^{n} \frac{(-1)^n n!(M-k-1)!}{(n-k)!M!} \sum_{i=0}^{k} \frac{(-1)^i M!}{(M-i)!i!} \\
&= \frac{1}{n!} \sum_{k=0}^{n} \frac{(-1)^n \binom{n}{k}}{\binom{M-1}{k}} \sum_{i=0}^{k} (-1)^i \binom{M}{i} \\
&= \frac{1}{n!} \sum_{k=0}^{n} \frac{(-1)^n \binom{n}{k}}{\binom{M-1}{k}} (-1)^k \binom{M-1}{k} \\
&= \frac{(-1)^n}{n!} \sum_{k=0}^{n} (-1)^k \binom{n}{k} = \begin{cases} 1, & n=0 \\ 0, & 0 < n \leq M-1, \end{cases}
\end{aligned}$$

where we use the equality [13]: $\sum_{i=0}^{k} (-1)^i \binom{M}{i} = (-1)^k \binom{M-1}{k}$. With a similar approach, we can obtain $a_M = -\dfrac{1}{M!}$, therefore

$$P_1(x) = 1 - \frac{1}{M!} x^M + o(x^M). \tag{71}$$

On the other hand, $P_2(x)$ can be represented as

$$P_2(x) = Me^{-x} \sum_{k=M+1}^{M+N_T-2} \left\{ \frac{(k-M-1)!}{k!} \left[ \sum_{i=0}^{k-M-1} \frac{x^{i+M}}{i!} \right] \right\},$$

and after some manipulations we obtain

$$P_2(x) = b_M x^M + o(x^M), \tag{72}$$

where $b_M = M \sum_{k=0}^{N_T-3} \frac{k!}{(M+k+1)!}$. By combining (70), (71) and (72), we can derive (33). Furthermore, from [13], we have

$$\sum_{k=1}^{\infty} \frac{k!}{(M+k+1)!} = \frac{1}{M(M+1)!},$$

therefore

$$\sum_{k=0}^{N_T-3} \frac{k!}{(M+k+1)!} < \sum_{k=0}^{\infty} \frac{k!}{(M+k+1)!}$$
$$= \sum_{k=1}^{\infty} \frac{k!}{(M+k+1)!} + \frac{1}{(M+1)!} = \frac{1}{M(M+1)!} + \frac{1}{(M+1)!} = \frac{1}{M.M!},$$

so the coefficient of $x^M$ in (33) is always positive.

[11]   D. Pedoe, *A Course of Geometry for Colleges and Universities*, Cambridge: University Press, 1970.

[12]   S. Loyka, and F. Gagnon, "Performance analysis of the V-BLAST algorithm: an analytical approach, " *IEEE Trans. Wireless Communications*, vol. 3, no. 4, pp. 1326-1337, July 2004.

[13]   I. S. Gradsbteyn, and I. M. Ryzbik, *Table of Integrals, Series, and Products*, 6$^{th}$ Edition, Academic Press, 2000.

[14]   M. Gharavi-Alkhansari, and A. B. Gershman, "Fast antenna subset selection in MIMO systems", *IEEE Trans. Signal Processing*, vol. 52, no. 2, pp. 339-347, Feb. 2004.

[15]   H. Zhang, and H. Dai, "Fast transmit antenna selection algorithms for MIMO systems with fading correlation," *Proc. Vehicular Technology Conference, Fall 2004, VTC Fall 04*, Sept. 2004.

[16]   Z. Wang, and G. B. Giannakis, "A simple and general parameterization quantifying performance in fading channels," *IEEE Trans. Communications,* vol. 51, no. 8, pp. 1389-1397, Aug. 2003.

[17]   N. Prasad and M. K. Varanasi, "Analysis of decision feedback detection for MIMO Rayleigh fading channels and optimization of power and rate allocations," *IEEE Trans. Inform. Theory*, vol. 50, no. 6, pp. 1009-1025, June 2004.

[18]   R. J. Muirhead, *Aspect of Multivariate Statistics Theory*, New York: Wiley, 1982.

[19]   B. Hassibi, "Random matrices, integrals and space-time systems," *DIMACS Workshop on Algebraic Coding and Information Theory*, Rutgers University, Dec. 2003.

[20]   H. A. David, Order Statistics, 3$^{rd}$ edition, John Wiley & Sons, 2003.

[21]   C. D. Meyer, *Matrix analysis and applied linear algebra*, SIAM, 2000.
25

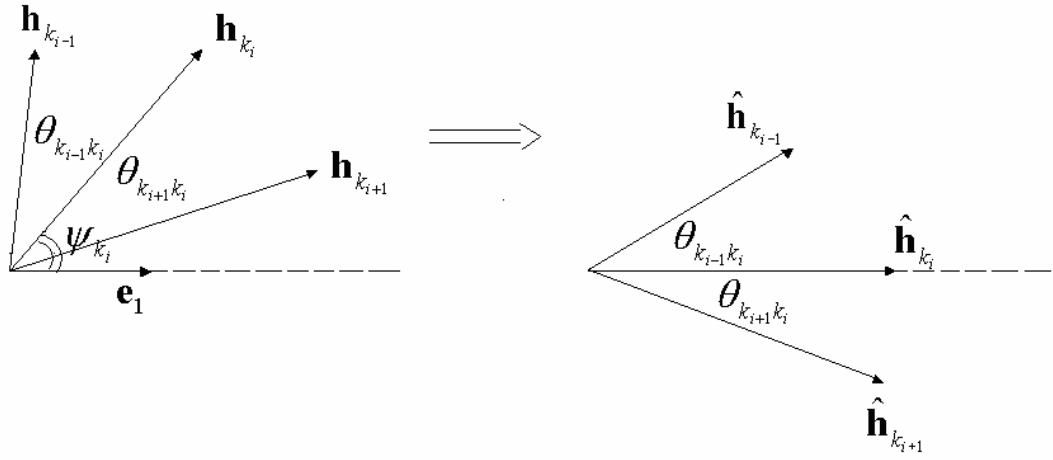

*Figure 1. Independence of $\theta_{k_{i-1}k_i}$ and $\theta_{k_{i+1}k_i}$*

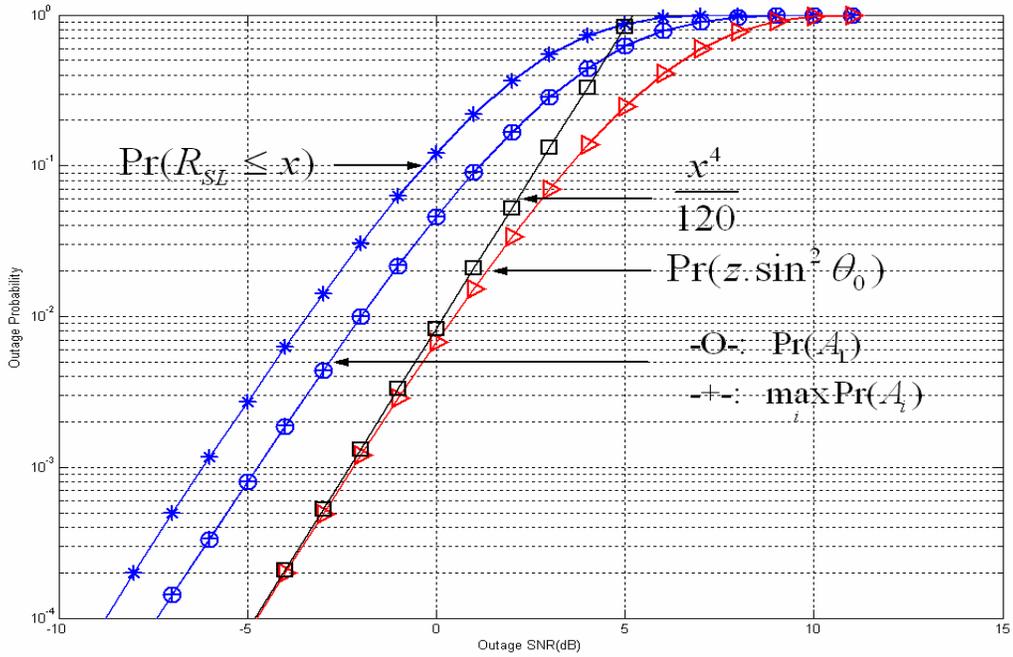

*Figure 2. The Exponential Behavior of $\Pr(A_1)$ for the $N_T = 3$, $N_R = 3, L = 2$ Scenario*

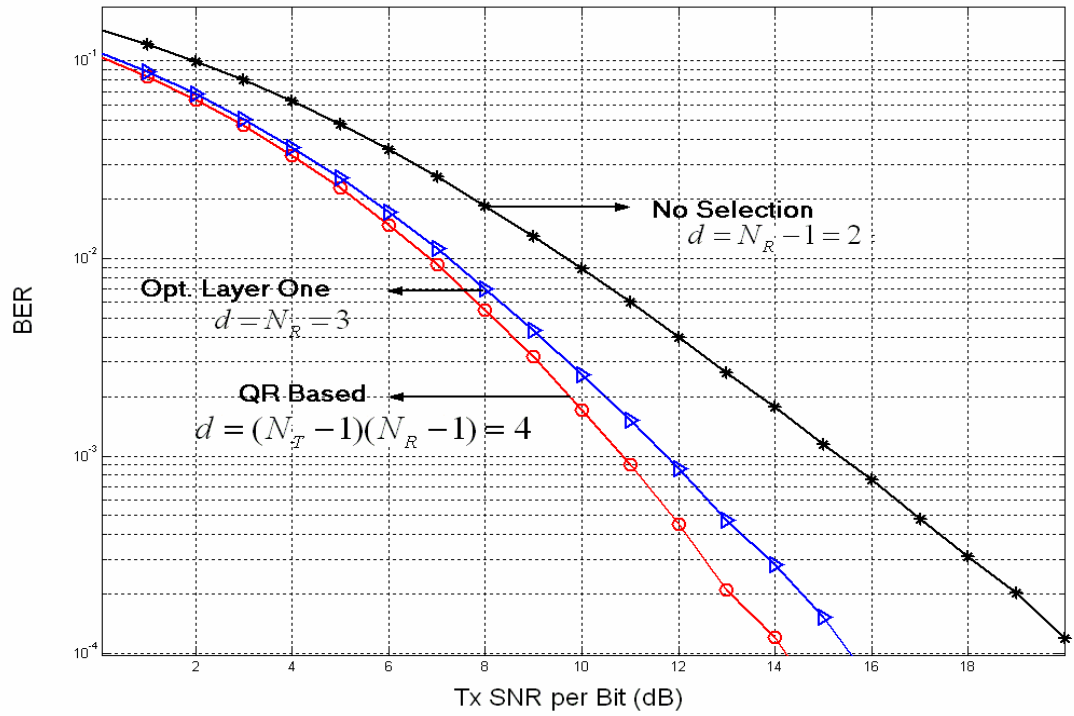

*Figure 3. BER Performance of SIC Receiver with* $N_T = 3$, $N_R = 3$, $L = 2$